# Characterization and Predictive Modeling of Epitaxial Silicon-Germanium Thermistor Layers

B. Gunnar Malm, Mohammadreza Kolahdouz, Fredrik Forsberg, Frank Niklaus
School of ICT and EES
KTH Royal Institute of Technology
Kista, Sweden
gunta@kth.se

***Abstract*—The thermal coefficient of resistance (TCR) for epitaxial silicon-germanium (SiGe) layers has been analyzed by experiment and simulation. Predictive simulation using drift-diffusion formalism and self-consistent quantum-mechanical solutions yielded similar results, TCR around 2%/K at 300 K. This modeling approach can be used for different, graded and constant, SiGe profiles,. It is also capable of predicting the influence of background auto-doping on the TCR of the detectors**

## I. Introduction

Semiconductor based thermistor materials are suitable for use in both cooled and uncooled thermal infrared bolometers [1]. For the integration with CMOS-based readout integrated circuits (ROIC) low-temperature deposition on pre-processed circuits or alternatively wafer bonding techniques are viable. The thermal isolation of active bolometer pixels is typically achieved by implementing suspended and thermally isolated membrane structures [2]. Thermistor materials based on alternating silicon (Si) and silicon-germanium (SiGe) epitaxial layers have been demonstrated and their performance is continuously increasing [3]. Compared to a single layer of silicon or SiGe, the temperature coefficient of resistance (TCR) can be strongly enhanced, by using thin alternating layers, which exhibit negative temperature dependence. In comparison it should be noted that doped silicon exhibits a weak positive TCR of less than 0.5 %/K. It has been observed that the TCR values in alternating Si/SiGe layers, sometimes referred to as quantum wells, are exponentially related to the bandgap offset between the silicon and SiGe layers. The bandgap offset causes a redistribution of charge carriers in the structure, resulting in a higher resistance as compared to a uniform carrier profile in a layer with the same temperature dependent mobility. The TCR value depends on the carrier concentration difference at the hetero-junctions, which is strongly dependent on the temperature. Previous modeling results were based on a numerical solution of the Poisson and carrier continuity equation in the drift-diffusion approximation, using a standard semiconductor device simulator [4]. Appropriate and reliable models for strained silicon-germanium are well proven in the context of strained silicon and SiGe devices [5]. In this work we have designed samples with a simplified layer structure to facilitate easier fabrication and to highlight important concepts in the modeling. The influence of quantum-mechanical effects, such as carrier confinement and tunneling, has been incorporated by self-consistently solving the 1-dimensional Schrödinger equation in the relevant parts of the layer structure, i.e. the vicinity of the hetero-junctions. An intermediate approximation level, the so called density gradient method (DG) [6], has also been shown to yield very similar results.

## II. Experimental Details

Three prototypes were grown using a RPCVD reactor on 200 mm SOI wafers. The SOI Si layer thickness was 70 nm on a 146 nm buried oxide. The devices include a single SiGe layer, sandwiched between two intrinsic spacer layers and highly boron-doped silicon layers from top and bottom, as shown in Fig.1. The role of the intrinsic spacer layer is to reduce auto-doping of the SiGe layer from the highly boron-doped contact layers, but also to tune the resistance of the layers, see below. Square pixels, of typical size for a bolometer array, are then formed through a lithography and dry-etching step in different sizes (200×200, 100×100, 50×50 and 25×25 µm2). $CF_4$, HBr, and $Cl_2$ gases were used to form the mesa by etching the epitaxially grown layers. Dry etching was carefully stopped at the center of the bottom boron-doped layer to obtain an ohmic contact.

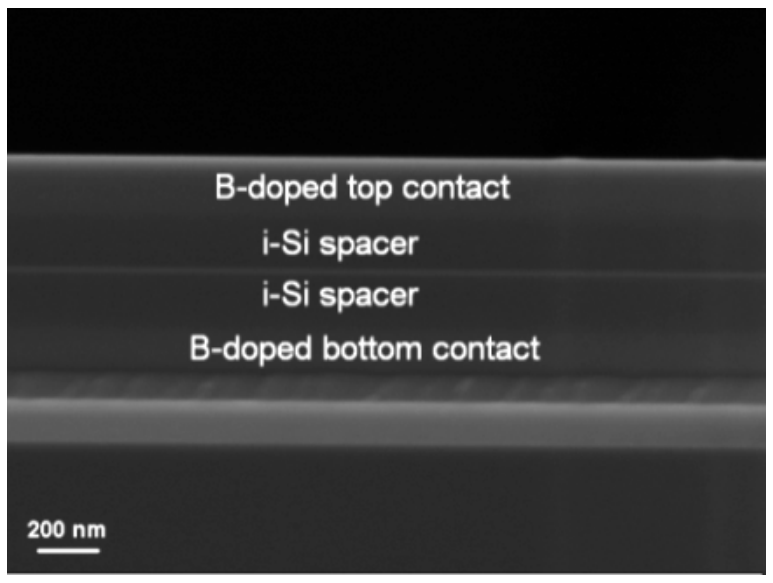

Figure 1. High resolution cross-sectional scanning electron microscopy of the grown layer for this study.

This work was partly supported by VINNOVA, Swedish Governmental Agency for Innovation Systems

After standard oxide passivation and etching the contact holes, Ni silicidation was performed. The process flow was ended by a TiW/Al metallization and a forming gas anneal (FGA) treatment. The SiGe layer thickness was 15 nm for all samples. The Ge content of the layer in these structures is 30% and 0.2 % of carbon was used to increase strain stability and to reduce boron diffusion. A sample was also designed with a graded Ge content from 0 to 30% (top to bottom), similar to a SiGe base layer in a HBT. To study the carrier distribution effect on the TCR, SiGe layers with and without a boron-spike were included, as shown in Fig. 2. Auto-doping during epitaxy results in background auto-doping in the spacer layers, surrounding the SiGe layer. We estimate the boron doping level to be in the range of $1\times10^{16}$ $cm^{-3}$, slightly below the resolution of the noisy SIMS profile.

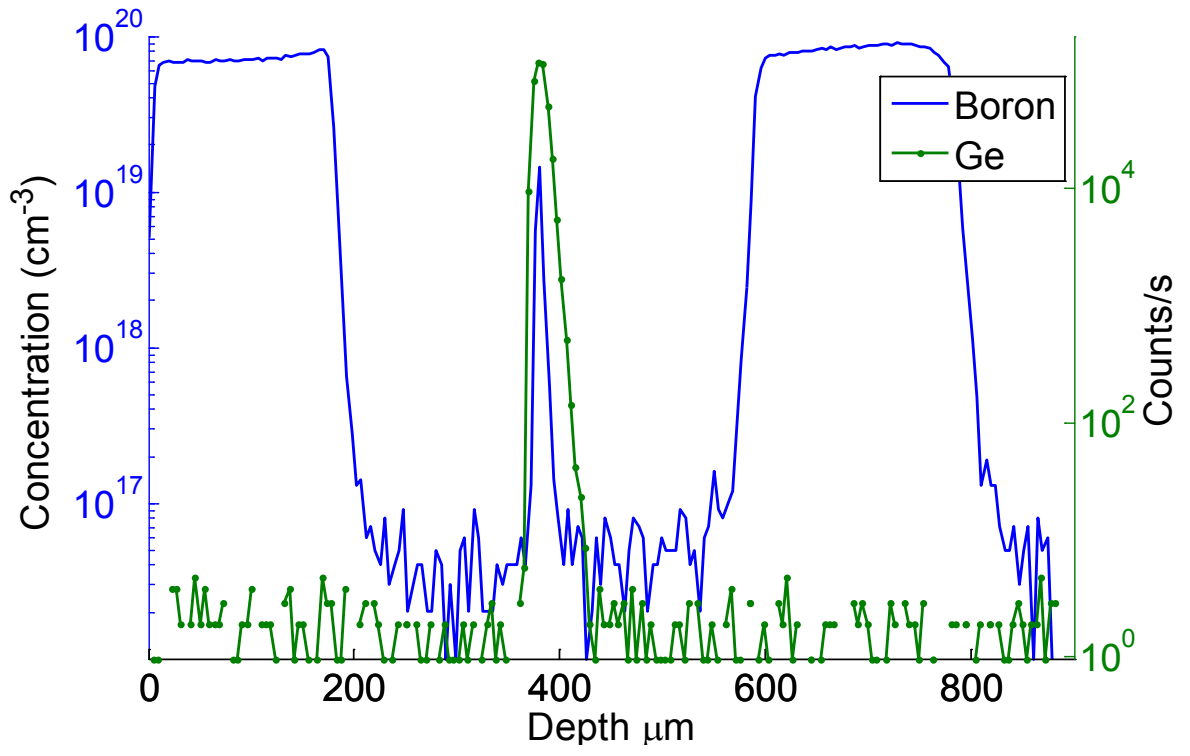

Figure 2. SIMS of box-layer type SiGe-layer (peak concentration 30 %) with stabilizing boron spike. Samples with graded 0-30 % SiGe layer and without a spike were also fabricated.

## III. Initial Experimental Results

Samples were characterized by current-voltage (I-V ) measurements over a temperature range of approximately 25 to 70 °C. The resistance shows a pronounced temperature dependence i.e. high negative TCR, as shown in Fig. 3. There is also slight bias dependence of the resistance. Therefore Fig. 4 illustrates resistance in an Arrhenius plot, at negative, positive and close to zero bias (0.04 V). Samples, with or without the extra boron spike, are compared and they can both be satisfactorily fitted over the temperature range of interest with an expression of the type:

$$R(T) = R_0 e^{\frac{E_A}{kT}} \qquad (1)$$

Where $k$ is Boltzmann's constant. Taking the derivative we get the TCR as a function of temperature:

$$TCR = \frac{1}{R(T)}\frac{\partial R}{\partial T} = -\frac{E_A}{kT^2}(\%/K) \qquad (2)$$

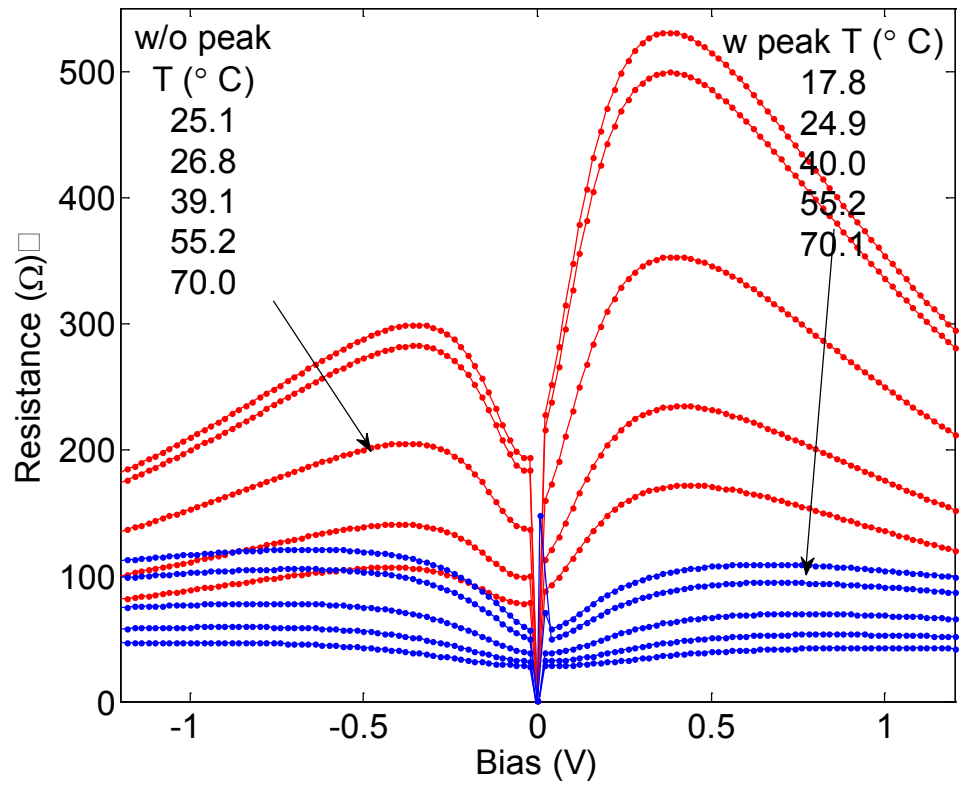

Figure 3. Resistance vs. bias and temperature, pixel size 50×50 μm$^2$. Family of IV-curves with lower values corresponds to boron peak.

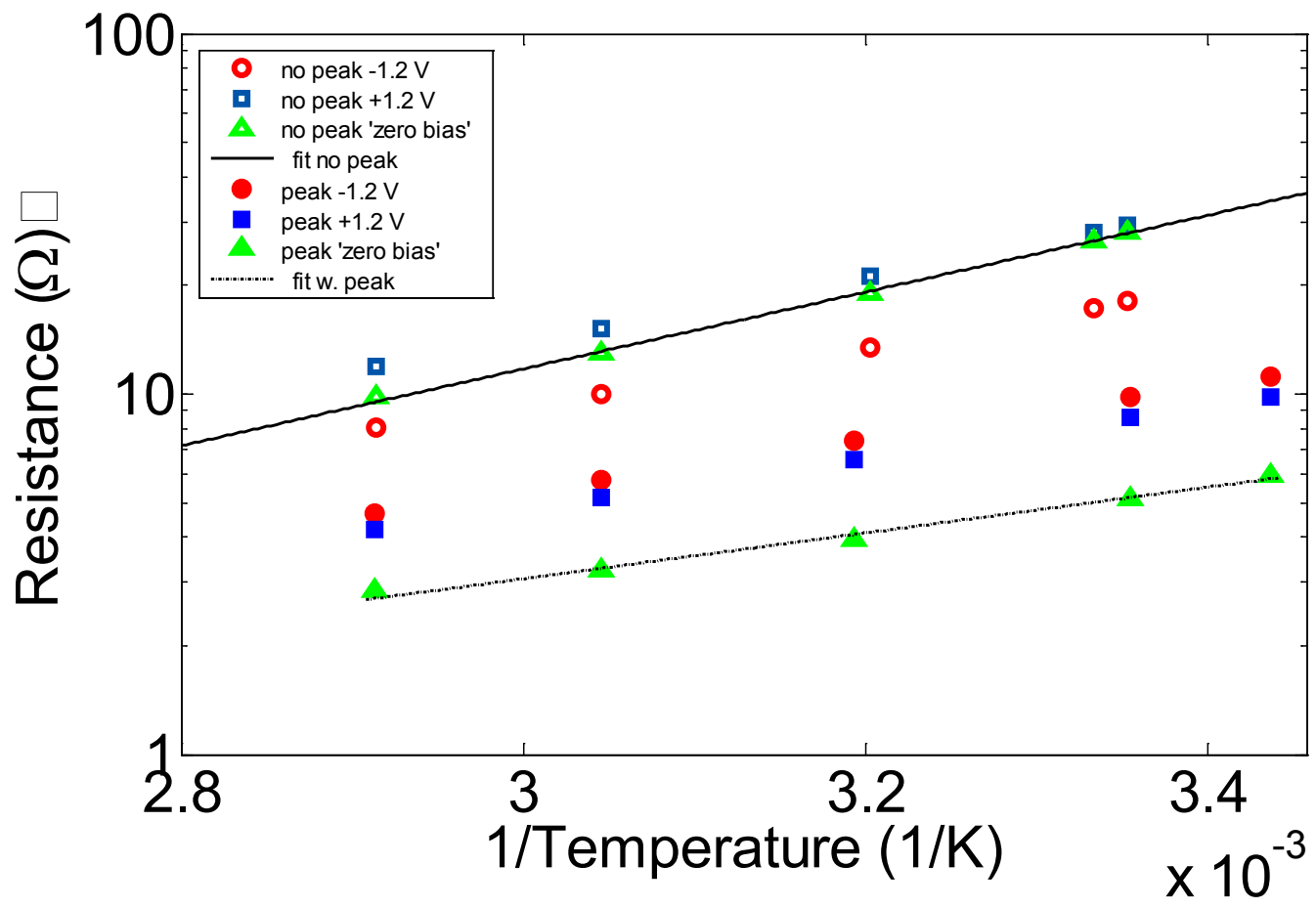

Figure 4. Resistance with and without boron peak vs. temperature at at negative, positive and close to zero bias (0.04 V).

The extracted TCR at 25 and 70 °C values are summarized in Table 1. Reliable values could not be obtained at 0 V. Therefore values at 0.04 V are reported. It can be observed that the two types of samples have TCR in the same range and also that the boron peak reduces the resistance significantly while the relative influence on the TCR is smaller. This result suggests that tuning of the doping, to get a resistance that is suitable for practical bolometer applications is feasible, since the TCR remains in a useful range.

TABLE I. Extracted TCR from Arrhenius plots

| Bias (V) | No peak (%/K) | | With peak (%/K) | |
|---|---|---|---|---|
| | *25 °C* | *70 °C* | *25 °C* | *70 °C* |
| -1.2 | -2.1 | -1.6 | -1.9 | -1.4 |
| 0.04 | -2.8 | -2.1 | -1.6 | -1.2 |
| +1.2 | -2.3 | -1.8 | -1.9 | -1.4 |

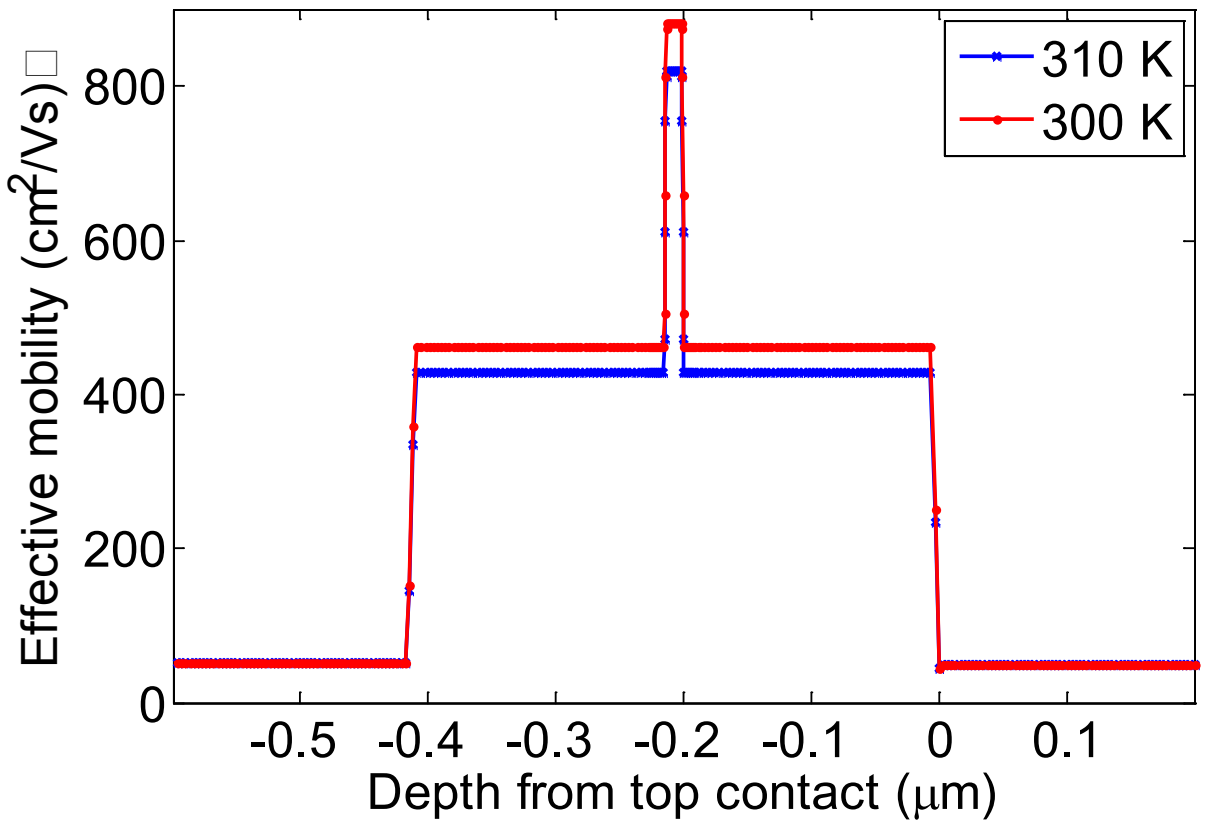

Figure 5. Temperature dependent mobility extracted from the device simulator

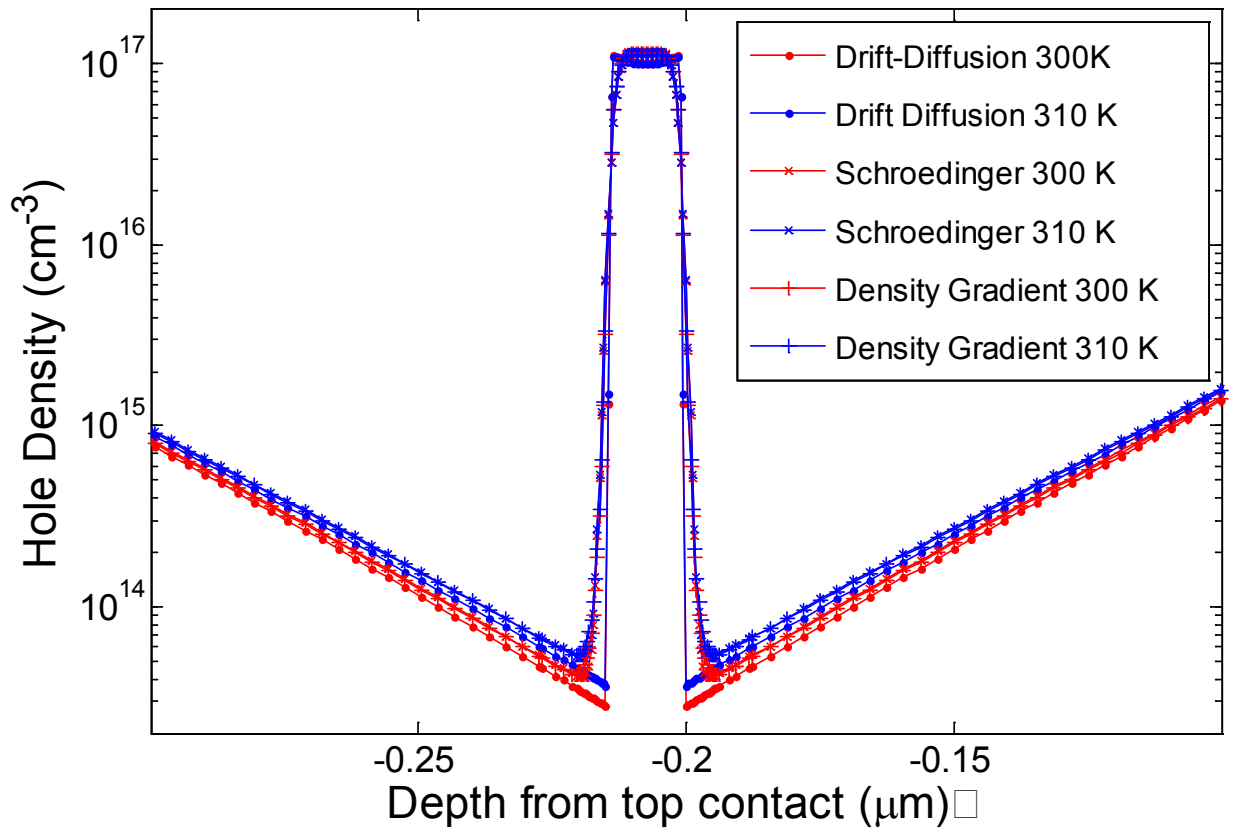

Figure 6. Zoomed-in view of hole concentration in the i-Si and SiGe regions, for classical and quantum-mechanical cases.

## IV. Modeling Classical and Quantum Mechanical Approach

In order to calculate the TCR the I-V characteristics of different structures were simulated as function of temperature using the Synopsys Sentaurus TCAD tool [7]. The total resistance ($R_{TOT}$) can be extracted from the bias dependent slope of the I-V curve or alternatively by direct integration of the position dependent resistivity for a specific applied bias.

$$Rtot \times area = \int_0^L \rho(x)\, dx = \int_0^L \frac{1}{qp(x)\mu_p(x)} dx \quad (3)$$

Obviously these two methods yield identical results. It is also clear that a region with a small carrier concentration $p(x)$ will result in a larger relative contribution, to $R_{TOT}$, if the mobility variation $\mu_p(x)$ in the whole structure is small. To highlight the mechanism of a strong negative TCR we perform the $R_{TOT}$ calculations from the local carrier density, essentially the hole profile, since the electron concentration can be neglected in this case, where all of the structure is boron (p-type) doped. Also needed is the effective mobility as calculated by the simulator, including the temperature, field (velocity saturation) and doping dependence. Figures 5 and 6 shows the extracted effective hole mobility and hole carrier concentration vs. depth, for zero applied bias at 300 and 310 K respectively with a 15 nm box shaped 30% SiGe layer placed symmetrically between highly doped top and bottom contacts. Note that the mobility is reduced for higher temperature. This is the normal behavior for a doped semiconductor layer. Nevertheless the TCR will turn out to be negative, since the carrier concentration plays a much bigger role in the calculation. The results for the drift-diffusion, quantum-mechanical (Schrödinger equation) and density gradient solutions are essentially overlapping, as shown by the zoomed-in view of the hole concentration in the intrinsic Si spacer and SiGe regions in Fig. 6.

The two quantum-mechanical solutions actually demonstrate a smoother transition in carrier concentration at the location of the hetero-junction, while the ratio of carrier concentration, inside and outside the SiGe layer, is almost unchanged , along with the TCR value.

As shown in Fig. 6 the presence of the hetero-junction induces an elevated carrier concentration inside the SiGe layer. The increased concentration is needed to accommodate the potential difference, caused by the bandgap offset. At the same time, the concentration outside the SiGe layer is reduced below the background doping level of $1\times10^{16}$ cm$^{-3}$, in order to maintain the total net charge, given by the acceptor dopant profile. The resulting effect is that the silicon low doped regions exhibit large resistivity, while the top and contact region and the SiGe layer itself will have little contribution to the total resistance of the structure. If one plots the resistivity at 310 K normalized to 300 K the origin of the effective negative TCR is easily seen. To highlight the difference between a Si/SiGe heterojunction structure and a silicon layer with the same dopant profile this case is also included in the plot, as indicated in Fig. 7. The calculated TCR values are -1.53 %/K and +0.46 %/K respectively. Using either the Schrödinger equation or the DG formalism we obtain -1.38 and -1.37 %/K for a SiGe layer, while a silicon layer is not affected by quantum-mechanical corrections.

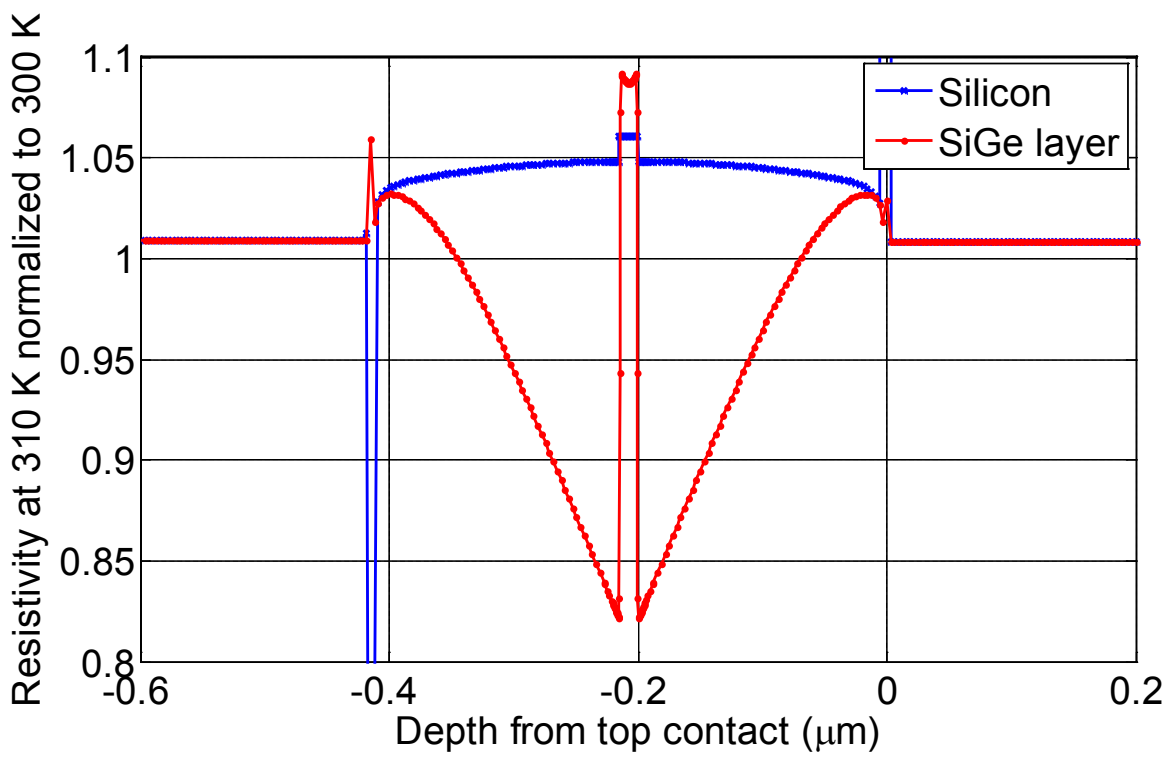

Figure 7. Origin of negative TCR for a sample with SiGe layer, as compared to doped silicon with positive TCR.

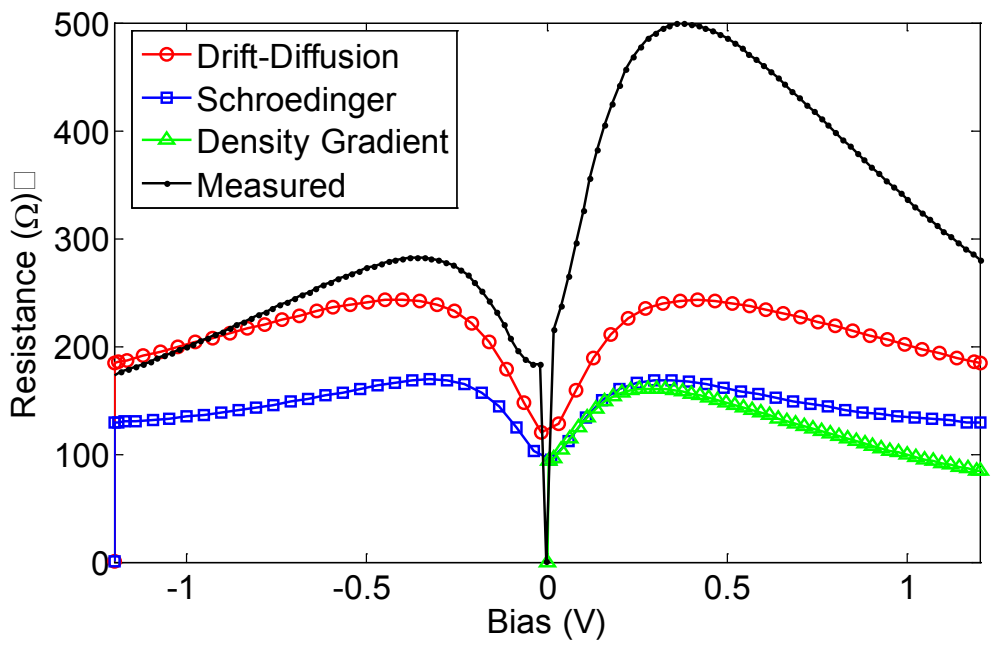

Figure 8. Comparison of simulated and measured resistance at 300 K vs. bias for different models, pixel size 50×50 μm$^2$

## V. PREDICTIVE CAPABILITY

A more detailed comparison of the different simulation approximation levels shows that the agreement is best at low bias/low field, see Fig. 8. The calculated resistance values agree well at low bias, while at higher bias the quantum-mechanical and DG curves are close while the drift-diffusion predicts higher resistance. This is in good agreement with the carrier profiles shown above, where more abrupt gradients in the vicinity of the heterojunction result in higher $R_{TOT}$. The measured data show an asymmetry for this particular sample but the overall predictive capability is convincing.

One unknown parameter in the simulations is the auto-doping level in the intrinsic silicon regions. Initially $1\times10^{16}$ cm$^{-3}$ was used, but to examine the auto-doping a range of $1\times10^{12}$ cm$^{-3}$ to $1\times10^{17}$ cm$^{-3}$ was considered. The results are illustrated in Fig. 9 and show that the influence of doping is significant for a concentration higher than $1\times10^{15}$ cm$^{-3}$.

Finally, we demonstrate the TCR vs. bias for a graded SiGe profile. From the discussion above it can be understood that the carrier concentration under forward and reverse bias in such a structure would be different. This could be translated into a strongly bias dependent resistance and TCR. Both experimental results and simulations confirm this, as illustrated in Fig 10. While the qualitative agreement between measurement and simulation is still satisfactory this case is more challenging, than the box-type profile discussed above. A SIMS profile of the graded structure was not available so the simulated and measured devices could have different grading as well as a different maximum concentration.

## VI. CONCLUSIONS

Good predictive capability for the simulation of TCR values in alternating Si/SiGe epitaxial layers is demonstrated. Influence of auto-doping level, boron peak and graded Ge-profile could be studied. A detailed comparison of drift-diffusion and quantum-mechanical simulations was presented and results were discussed in terms of carrier profile influence on resistivity vs. depth in the thermistor material

## ACKNOWLEDGMENT

We are grateful to C-M Zetterling, P-E Hellström, and V. Guðmundsson for helpful discussions.

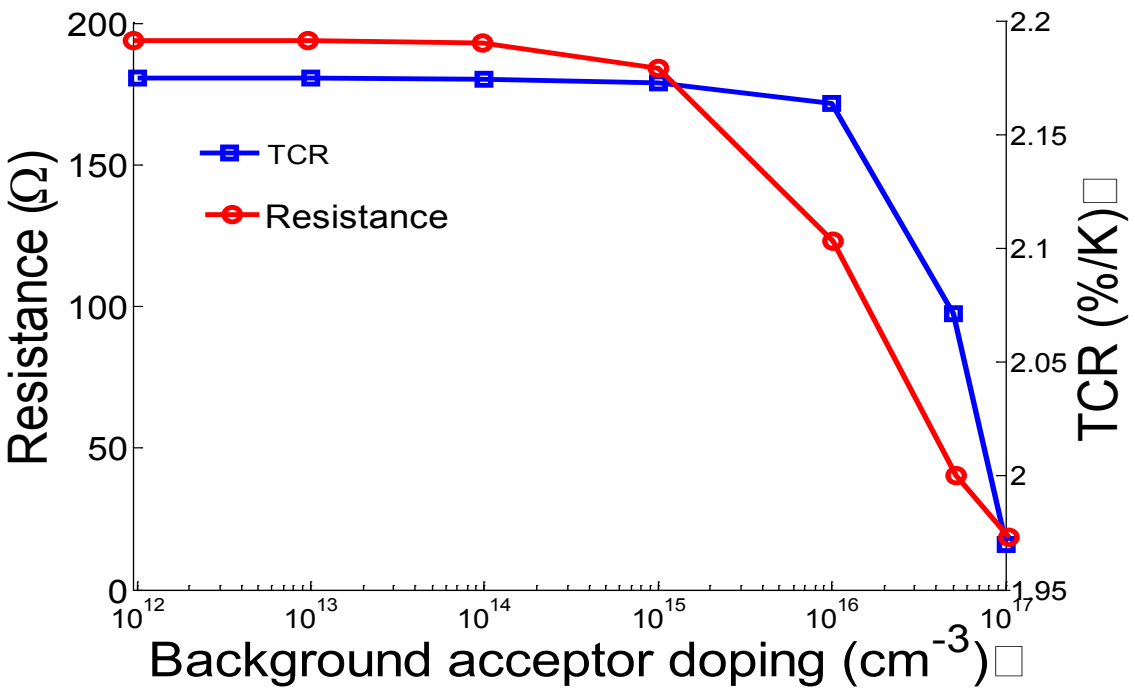

Figure 9. Simulated total resistance and TCR variation with and background doping at +1.2 V bias

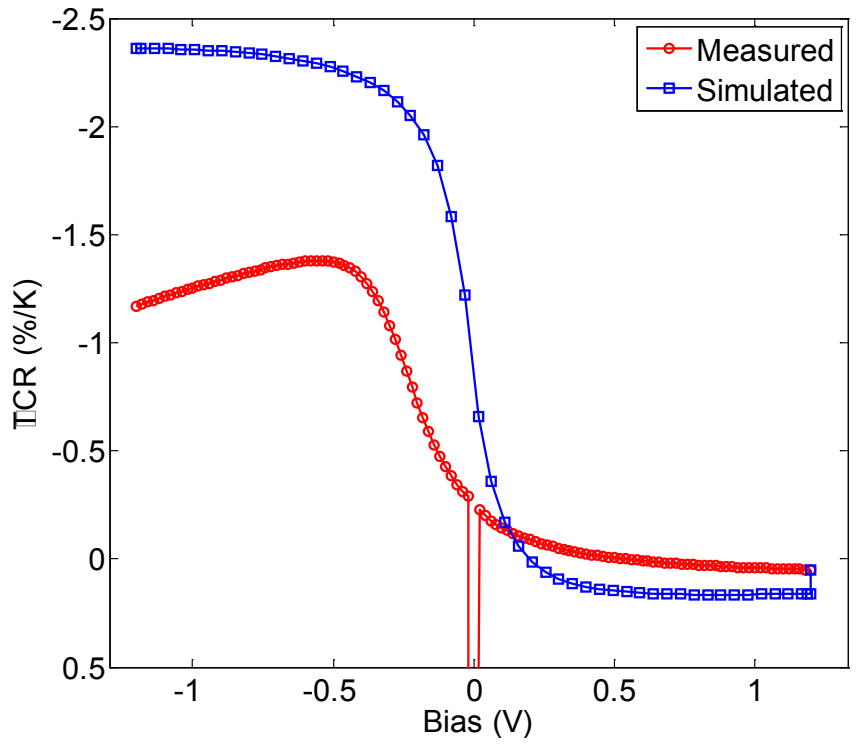

Figure 10. Simulated and measured TCR for a graded SiGe profile